\documentclass[superscriptaddress,hyperref,aps,pra,twocolumn,floatfix,nofootinbib]{revtex4-2}

\usepackage{graphicx,xcolor,soul}
\usepackage{bm}
\usepackage{dsfont}
\usepackage{amsmath, amsthm, amssymb,mathrsfs,dsfont}
\usepackage{setspace}
\usepackage[utf8]{inputenc}
\usepackage[toc,page]{appendix}
\usepackage{hyperref}
\usepackage{mdframed}
\usepackage{footnote}
\usepackage{fancyhdr}

\newcommand{\ket}[1]{\left| #1 \right\rangle}
\newcommand{\bra}[1]{\left\langle #1 \right|}
\newcommand{\proj}[1]{\ket{#1}\hskip-1mm\bra{#1}}
\newcommand{\av}[1]{\langle #1\rangle}
\newcommand{\braket}[2]{\langle #1|#2 \rangle}
\newcommand{\ketbra}[2]{\left|#1\right\rangle\hskip-1mm\left\langle#2\right|}

\begin{document}
\title{Weak values and the past of a quantum particle}

\author{Jonte R. Hance}
\email{jonte.hance@bristol.ac.uk}
\affiliation{Quantum Engineering Technology Laboratories, Department of Electrical and Electronic Engineering, University of Bristol, Woodland Road, Bristol, BS8 1US, UK}
\author{John Rarity}
\affiliation{Quantum Engineering Technology Laboratories, Department of Electrical and Electronic Engineering, University of Bristol, Woodland Road, Bristol, BS8 1US, UK}
\author{James Ladyman}
\email{james.ladyman@bristol.ac.uk}
\affiliation{Department of Philosophy, University of Bristol, Cotham House, Bristol, BS6 6JL, UK}

\begin{abstract}
    We investigate four key issues with using a nonzero weak value of the spatial projection operator to infer the past path of an individual quantum particle. First, we note that weak measurements disturb a system, so any approach relying on such a perturbation to determine the location of a quantum particle describes the state of a disturbed system, not that of a hypothetical undisturbed system. Secondly, even assuming no disturbance, there is no reason to associate the non-zero weak value of an operator containing the spatial projection operator with the classical idea of `particle presence'. Thirdly, weak values are only measurable over ensembles, and so to infer properties of individual particles from values of them is problematic. Finally, weak value approaches to the path of a particle do not provide information beyond standard quantum mechanics (and the classical modes supporting the experiment). We know of no experiment with testable consequences that demonstrates a connection between particle presence and weak values. 
\end{abstract}

\maketitle

\section{Introduction}

Weak values are obtained via weak measurement and pre- and post-selection, of sub-ensembles from ensembles of particles \mbox{\cite{Aharonov1988Weak}}. They have been shown to be very useful for metrology and state tomography tasks.\cite{Dixon2009WeakAmp,Jordan2014AdvWeak,Harris2017Weak}. Weak values have also been used to infer the presence of individual quantum particles, where standard quantum mechanics would not normally allow this. According to these weak value approaches to particle presence, non-zero weak values show where a quantum particle has been, even when this would normally be impossible due to the pre- and post-selected states not being eigenstates of spatial projection operators \mbox{\cite{Danan2013Asking,Vaidman2014Tracing}}. Despite weak values only being defined over ensembles, in these approaches, inferred particle presence is attributed to individual particles.

One such approach is the Weak Trace Approach, according to which particles leave traces along the paths they traverse, due to interaction (coupling) with the environment, and such a trace is left even when this interaction is insufficient to count as a projective measurement. A non-zero weak value of an operator on a path (weak trace) is taken to show that an individual particle travelled along that path \mbox{\cite{Vaidman2013Past}} even in the limit of this interaction (coupling) vanishing.

According to this approach there is a weak trace in a region when any operator formed by the product of the spatial projection operator for that region with an operator for a property of the particle (e.g. spin) has a non-zero weak value. This can happen even if the weak value of the spatial projection operator itself is zero. Weak traces are supposed to give us more information about the particle than von Neumann measurement of the spatial projection operator would allow. The Weak Trace Approach even allows discontinuous particle trajectories, hinting at a mechanism by which seemingly disconnected events can affect one another, such as in counterfactual communication \mbox{\cite{Salih2013Protocol}}. Therefore, this approach has been controversial \cite{Li2013CommVaidman,Vaidman2013LiPastReply,Englert2017Revisited,Griffiths2016Path,Vaidman2017CommentGriffith,Griffiths2017Reply,Salih2018CommentPath,Griffiths2018Reply,Hashmi2016TSVF,Vaidman2016PastInterferom}. 

Another approach connecting weak values and particle presence, implicit in the quantum Cheshire cat \mbox{\cite{Aharonov2013CheshireCats}}, requires that the spatial projection operator at a particular location has a non-zero weak value, for us to say the particle was present at that location. In the quantum Cheshire cat, these approaches come apart since the weak value of the spatial projection operator on a path can be zero, but the weak value of the spin operator times the spatial projection operator non-zero---in these cases, according to the Weak Trace Approach the particle \emph{was} on the path with a zero weak value for the spatial projection operator, contrary to \mbox{\cite{Aharonov2013CheshireCats}}.

In this paper, we show how one obtains the weak value of an operator. We then give a case where the weak value of an operator containing the spatial projection operator is unexpectedly non-zero (and so, by weak value approaches, a particle has left an unexpected trace in its environment). After this, we look specifically at the Weak Trace Approach. It is argued in \mbox{\cite{Peleg2018CommentRevisited}} that the Weak Trace Approach is justified because there is some non-vanishing local interaction between the quantum particle and its environment along whatever path it travels \mbox{\cite{Peleg2018CommentRevisited}}. This implies the weak value of the spatial projection operator along a path should always be non-zero whenever a particle travels along that path, and so the particle was not present if the weak value of the spatial projection operator is zero. However, there are cases where according to the Weak Trace Approach a particle is present on a path, despite the spatial projection operator for that path having weak value zero (e.g. in the quantum Cheshire cat set-up).

We then investigate four key issues with weak value approaches. 

First, even weak measurements disturb a system, so any approach relying on such a perturbation to determine the location of a quantum particle will only describe this disturbed system, not a hypothetical undisturbed state. We highlight this using the case of a balanced interferometer tuned to have destructive interference (i.e. no light exiting at its dark port) where such a perturbation changes the nature of the system. The unperturbed state is the vacuum, while the perturbed state has light present. While the measurement effect can be made arbitrarily small, this is not the same as removing it entirely. Further, we show attempts to respond to this line of argument by saying there are no completely unperturbed systems, as weak interactions are ubiquitous in nature (as in \mbox{\cite{Peleg2018CommentRevisited}}), raises questions about why weak value approaches only associate presence with the part of this interaction which is of the same order of magnitude as the coupling between system and environment.

Secondly, even assuming no disturbance, there is no reason to associate the non-zero weak value of an operator containing the spatial projection operator with the classical idea of `particle presence'. Indeed, in some situations, just taking the weak value gives features inconsistent with classical ideas associated with a particle being present (e.g., giving discontinuous particle trajectories, or particles not being in coarse-grainings of their location).

Thirdly, weak values are only measurable over ensembles, and so to infer properties of individual particles from values of them is problematic at best.

Finally, weak value approaches to the path of a particle do not give us any new physics beyond standard quantum theory \mbox{\cite{Vaidman1996WeakElReality}}, to explain the causes of counterintuitive quantum effects (or even the paradoxes the approach itself creates). They assume a connection between particle presence and weak values without contributing testable new physics.

\section{Weak Values}\label{WeakVals}

We first present the derivation of a weak value of an operator, and how it is argued this relates to the trace left by a particle on its environment.

The weak value $O_w$ of an operator $\hat{O}$ is defined as \cite{Aharonov1988Weak}
\begin{equation}\label{Eq:WeakVal}
    O_w=\av{\hat{O}}_w=\frac{\bra{\psi_f}\hat{O}\ket{\psi_i}}{\braket{\psi_f}{\psi_i}}
\end{equation}

As this weak value increases as $\braket{\psi_f}{\psi_i}$ goes to zero, weak value protocols are used in metrology to amplify signals from delicate results so they can be observed experimentally \cite{Dixon2009WeakAmp,Jordan2014AdvWeak,Harris2017Weak}. This is at the expense of postselection reducing success probability. Weak values however also lead to a range of paradoxes \cite{Aharonov2005Paradoxes}.

In practice, the weak value is measured experimentally by taking the postselected result of a weak measurement. To obtain this weak value, we first couple our initial system $\ket{\psi_i}$ to our initial pointer state, $\ket{\phi}$, by weakly measuring them with the probe Hamiltonian $\hat{H}=\lambda\hat{O}\otimes\hat{P_d}/T$ for small coupling constant $\lambda$ and state-probe interaction time $T$. This produces the state
\begin{equation}
\begin{split}
     \ket{\psi_w}&=e^{-\frac{i\hat{H}T}{\hbar}}\ket{\psi_i}\otimes\ket{\phi}\\
     &=e^{-\frac{i\lambda}{\hbar}\hat{O}\otimes\hat{P_d}}\ket{\psi_i}\otimes\ket{\phi}
\end{split}
\end{equation}
where $\hat{P}_d$ is the momentum of that pointer. We then strongly measure this weak-measured state $\ket{\psi_w}$ with the operator 
\begin{equation}
  \hat{F}_1=\ketbra{\psi_f}{\psi_f}\otimes\hat{I}_d  
\end{equation}

Assuming $\hat{P}_d$ has Gaussian distribution around 0 with low variance (so $\hat{X}_d$, the position of the pointer, has Gaussian distribution with high variance),
we can say
\begin{equation}\label{Eq:FirstOrderApprox}
    \begin{split}
        &e^{-\frac{i\lambda}{\hbar}\hat{O}\otimes\hat{P_d}}=\sum_{k=0}^\infty\big(-\frac{i\lambda}{\hbar}\hat{O}\otimes\hat{P_d}\big)^k/k!\\
        &\;=\mathds{1}-\frac{i\lambda}{\hbar}\hat{O}\otimes\hat{P_d}+\mathcal{O}(\lambda^2)\approx \mathds{1}-\frac{i\lambda}{\hbar}\hat{O}\otimes\hat{P_d}
    \end{split}
\end{equation}
so this strong measurement gives the result
\begin{equation}
\begin{split}
&\ketbra{\psi_f}{\psi_f}e^{-\frac{i\lambda}{\hbar}\hat{O}\otimes\hat{P_d}}\ket{\psi_i}\otimes\ket{\phi}\\
&\approx\ketbra{\psi_f}{\psi_f}(\mathds{1}-\frac{i\lambda}{\hbar}\hat{O}\otimes\hat{P_d})\ket{\psi_i}\otimes\ket{\phi}\\
&=\ket{\psi_f}\otimes\braket{\psi_f}{\psi_i}(\mathds{1}-\frac{i\lambda}{\hbar}O_w\hat{P}_d)\ket{\phi}\\
    &\approx\ket{\psi_f}\otimes\braket{\psi_f}{\psi_i}e^{-\frac{i\lambda}{\hbar}O_w\hat{P}_d}\ket{\phi}
\end{split}
\end{equation}

This entails the position of our initial pointer state $\ket{\phi}$, acting as our ``readout needle", shifts, with measurement of the pointer position giving a read-out value $(\langle x\rangle - a)$. For an initial Gaussian pointer state, this shift causes the average position to move from $\langle x \rangle$ to $(\langle x \rangle - \overline{a})$ due to the application of the effective operator $\mathds{1}-i\lambda O_w\hat{P}_d/\hbar$. $a$ is distributed over a wide range of values for many repeats, but the distribution of ``$a$"s  will be a Gaussian centred on (the real part of) $\lambda O_w$. Peculiarly, this weak value $O_w$ can be very far from any of the eigenvalues of $\hat{O}$, or even imaginary \cite{Aharonov1988Weak,Tamir2013Intro,Dressel2014UnderstandingWeak}. This is odd, given this weak value appears in exactly the place in the equation that an eigenvalue of $\hat{O}$ would for a Von Neumann measurement (where the variance of $\hat{X}_d$ on the measured state is vanishingly small). This led to the belief that the weak value $O_w$ represented some fundamental value of the operator $\hat{O}$ \textbf{between} measurements.

\begin{figure}
    \centering
    \includegraphics[width=\linewidth]{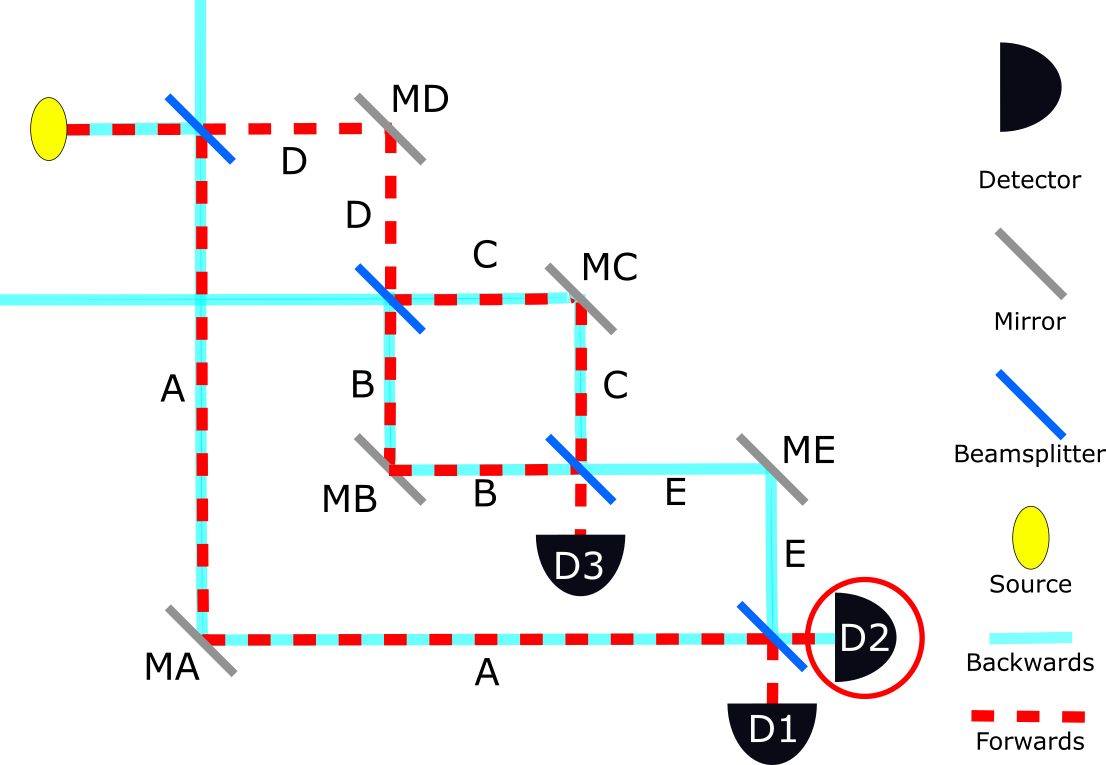}
    \caption{Nested-interferometer model, used to discuss cases where the Two-State Vector Formalism contradicts ``common-sense" continuous-path approaches \cite{Vaidman2013Past}.}
    \label{fig:TSVFDiag}
\end{figure}

Due to their interrelatedness, we can graphically show where the spatial projection operator has a non-zero weak value through the Two-State Vector Formalism (TSVF). The TSVF considers both the backwards and forwards-evolving quantum states, rather than just the forward as in standard quantum mechanics \cite{Aharonov1964TSVF}. (A similar intuition led to the development of the Double Inferential-state Vector Formalism \cite{Watanabe1955DIVF}.) For some operator $\hat{O}$, the forwards travelling initial state $\ket{\psi_i}$, and the backwards travelling final state $\bra{\psi_f}$, the TSVF gives out a conditional probability amplitude $\bra{\psi_f}\hat{O}\ket{\psi_i}$, where $\bra{\psi_f}\ket{\psi_i}$ is referred to as the Two-State Vector. This conditional probability amplitude is of the same form as the numerator of the weak value, and so we can use the TSVF to graphically plot where the weak value of an operator containing the spatial projection operator is nonzero. This involves plotting the forward-evolving state (possible paths the particle could have travelled via from its original position) and the backwards-evolving state (possible locations the particle could have come from to reach its final position). According to Weak value approaches, a quantum particle is present wherever these states visibly overlap (as shown in Fig. \ref{fig:TSVFDiag}).

The nested-interferometer model, which we show in Fig. \ref{fig:TSVFDiag}, was pointed out as a case where the Two-State Vector Formalism contradicted ``common-sense" continuous-path approaches \cite{Vaidman2013Past}. The inner interferometer is balanced such that a photon entering from arm $D$ exits to detector $D3$. Consequently, the outer interferometer is unbalanced, so there is an equal probability of a photon introduced from the source ending in $D1$ or $D2$. When the photon ends at $D2$, common sense would tell you it must have travelled via path $A$. This is as, had it travelled via $D$ into the inner interferometer, it could not have exited onto path $E$, or reached $D2$. However, according to the Weak Trace Approach, while the photon never travelled paths $D$ or $E$, it travelled along paths $B$ and $C$ as well as along path $A$.

\section{The Weak Trace Approach}\label{Vaidman}

Consider a case where the weak value of either the spatial projection operator, or an operator formed by multiplying the spatial projection operator with the operator for some property of a particle, along a certain path was non-zero for a given $\ket{\psi_i}$ and $\bra{\psi_f}$. According to the Weak Trace Approach this indicates a particle was on that path and left a weak trace along it \cite{Vaidman2013Past}. This is on two grounds. First, if it were a strong measurement, an operator containing the spatial projection operator having non-zero eigenvalues would be sufficient for a quantum particle to be present, so correspondingly a non-zero weak value of this operator is a weak trace of particle presence. Secondly, if a particle was present in a region, it should have non-zero interaction with the local environment (as reflected by the subtle decoherence its state would undergo), and so leave a ``trace'' along its local path. While this interaction is weak, which it must be so that the environmental coupling does not collapse position superpositions in the way a projective measurement would, according to the Weak Trace Approach it is still in principle detectable (i.e., the interaction would be of the same order of magnitude as $\lambda$), and so the interaction would be equivalent in scale to that required to give a non-zero weak value of an operator containing the spatial projection operator along that path.

However, there are cases where the Weak Trace Approach entails that a particle was present on a path, despite the spatial projection operator for that path having weak value zero. In these cases, the Weak Trace Approach takes it that a particle is present along a path even when the weak value of the spatial projection operator along that path is zero, so long as the weak value of some other operator multiplied by the spatial projection operator is nonzero (e.g. the weak value of the spin along the right-hand path in the quantum Cheshire cat set-up \cite{Aharonov2013CheshireCats}) \cite{Vaidman2013Past}. If the justification for the Weak Trace Approach is the non-vanishing local interaction, as in \cite{Peleg2018CommentRevisited}, then, if a quantum particle travels along a path, either: the weak value of the spatial projection operator on that path should be non-zero; or we should consider the higher-order terms in $\lambda$, as they would indicate additional information about the path on which the particle travelled; or we should say a non-zero weak value for the spatial projection operator is sufficient, but not necessary, to indicate the presence of a quantum particle, as this non-vanishing local interaction can exist along a path without there necessarily being a non-zero weak value for this operator. Hence, the Weak Trace Approach implicitly requires this third option, which raises the question of how it can be used to infer that particles were never present in some specific regions (see the nested interferometer, below). However, it could be that unlike the specific case of the weak value of the spatial projection operator, there existing at least one operator formed of the product of the spatial projection operator for a location and some other operator, with a non-zero weak value, is both a necessary and a sufficient condition for particle presence at that location. Further, it could be that in the nested interferometer case, this condition cannot be met at certain points, so the particle was not there according to the Weak Trace Approach.

While such an argument resolves any formal contradiction in associating a non-zero weak value for the product of some operator with the spatial projection operator with presence, even when the weak value of the spatial projection operator is zero, this still seems peculiar. A key part of the motivation behind weak value approaches is some assumption that the weak value of an operator tells us something about the property associated with that operator, similarly to an eigenvalue obtained by measuring that operator. By this assumption, while measuring the weak value of a spatial projection operator infers some information about whether the particle was at the location corresponding to that operator, measuring the weak value of some product of the spatial projection operator with some other operator corresponding to a property of the particle only infers some information about that property at that location. It requires the separate assumption that a property of a particle cannot be disembodied from that particle to then use the nonzero weak value of such a product operator to then infer about the particle's location.

\section{Issues with the Weakness Assumption}

The process of weak measurement and strong postselection given above leads to a small shift in the position $X_d$ of our pointer system $\ket{\phi(x)}$, with an average value
\begin{equation}
   \overline{a}=\lambda \text{Re}(O_w) << \Delta X_d
\end{equation}

Performing the measurement on a pre-and-postselected ensemble of $N$ particles allows measurement of the shift to precision $\Delta X_d/\sqrt{N}$. Therefore, according to the Weak Values Approach, as long as $N>(\Delta X_d)^2\mathcal{O}(\lambda^{-2})$, the presence of the particle is revealed. This is still however a coupling---so long as $\lambda\neq0$, there is measurement. This is as it must be, because Busch's Theorem says we cannot gain information about the state of the system without disturbing it \cite{Busch2009Limit}.

This is in tension with with the Weakness Assumption inherent to weak measurement---that if we measure weakly enough, we can effectively see how the system behaves when it is not measured. As argued above, weak interaction is still interaction and so disturbs the system (albeit negligibly so for some purposes). As long as there is a non-zero coupling, as in all weak value experiments, the `negligibly' small disturbance is still a disturbance, and there are differences in observable effects between small coupling and no coupling (see \cite{Hance2022DQCCDetectable} for an example of such a difference).

According to popular interpretations of weak values they exist in the absence of measurement and perturbation \cite{Aharonov2017Disappearing,Aharonov2018Mirage,Vaidman1996WeakElReality}. However, this is clearly not the case \cite{Svensson2013Weak,Wiesniak2014CriticismAsking,Salih2015CommentAsking,svensson2015non,Ben2017CommSvenNonRep,Kastner2017DemystifWeakVals,Svensson2017ReCommNonRep,Ipsen2022Disturbance,Hance2022DQCCDetectable}. To defend against this point \cite{Peleg2018CommentRevisited} explicitly abandons the Weakness Assumption, saying, ``in a hypothetical world with vanishing interaction of the photon with the environment, $[$the Weak Trace definition$]$ is not applicable, but in the real world there is always some non-vanishing local interaction. Unquestionably, there is an unavoidable interaction of the photon with the mirrors and beam splitters of the interferometer".

This statement, while formally correct and sufficient to avoid the issues associated with the Weakness Assumption, raises questions about why we should only consider the part of the weak trace which is of $\mathcal{O}(\lambda)$, especially when considered in the context of the approach taken practically in obtaining spatial weak values. If the purpose of weak value approaches is to identify the non-vanishing interaction a quantum particle has with its environment, to trace its path, then a definition of this trace which neglects some of these non-vanishing local interactions is only telling us part of the story. Weak value approaches only consider the first-order trace, as per the approximation given in Eq. \ref{Eq:FirstOrderApprox}, due to the supposedly vanishing nature of the terms of second-order or higher in $\lambda$. Despite this, \cite{Peleg2018CommentRevisited} specifically says that these environmental terms provide non-vanishing local interaction. A typical photon interaction with a mirror for instance transfers of order $10^{-33}$ of its energy per reflection (for a 1500nm-wavelength photon and a 1 gram mirror), a suitably weak interaction that is mostly safely ignored. Hence even the higher order terms are many orders of magnitude larger than the strongest perturbation from the environment in optics experiments, and so the invoking in \cite{Peleg2018CommentRevisited} of the weak interaction naturally left by the quantum object on its surroundings, while still having the weak trace ignore higher-order terms, seems more an attempt to avoid the issues with the Weakness Assumption than a physically well-motivated argument for believing the weak trace tells us everything about the location of a quantum particle.

It could be argued that this is more an objection to the experimental methods used to detect spatial weak values than the application of the concept itself---however, it is consideration of these experimental methods that is taken to motivate neglecting the higher-order terms (due to their comparative undetectability)---so therefore, if we are instead looking more theoretically at the non-vanishing local interactions, we should associate presence with all terms, rather than just the first order terms in $\lambda$.

The need to consider higher-order coefficients of $\lambda$ has also been discussed in \cite{Paneru2017PastEntg,aharonov2018completely,Georgiev2018SeqWeak,waegell2022quantum}.

\section{Do weak values reveal particle presence?}

In this Section, we argue that, even assuming no disturbance, there is no reason to associate the non-zero weak value of the spatial projection operator with the classical idea of particle presence. To make this argument, we first give a key assumption for this section---that presence is a property typically ascribed to classical particles, and so any attempt to form a definition of presence for quantum particles should correspond to our intuitions about classical presence, unless we have a good reason for it to deviate from this.

The classical conception of a particle presence---being present at a certain place at a certain time---can be characterised as follows:
\renewcommand{\labelenumi}{\roman{enumi}).}
\begin{enumerate}
    \item Every particle is located in space at all times.
    \item Particles cannot be on more than one path simultaneously.
    \item Particle trajectories are continuous (or at least as continuous as space is) so particles cannot get from one place to another without passing through the space in between.
    \item Particles interact with other objects/fields local to their location.
    \item If a particle is on a path at a given time, and that path is within some region, then the particle is also located in that region at that time.
    \item If a particle's property is at a location, the particle must be at that location too.
\end{enumerate}

In standard quantum mechanics, when a particle is in an eigenstate of a position operator, then it is attributed a position. However, quantum particles can be in superpositions of position eigenstates. Whether they then have positions at all, or even have two positions at once, is contentious and interpretation-dependent. To require that there is always some location at every time where conditions (i) and (ii) are satisfied is to advocate a hidden variable approach to quantum mechanics, where the hidden variable is the particle's location. However, despite quantum tunnelling and other phenomena, a reformulation of condition (iii) still applies in so far as the probability current of quantum particles evolving according to the Schr\"odinger equation obeys a continuity equation. (iv-vi) are also compatible with standard quantum mechanics.

Particles that are not in an eigenstate of one of the spatial projection operators demarcating coarse-grained locations of interest (e.g. specific paths in a nested-interferometer experiment) can nonetheless have nonzero weak values for any of those spatial projection operators (or any composite operators formed of any of those spatial projection operators multiplied by some other operator). Perhaps a condition for a particle being present at a location in a quantum context is that it has a non-zero weak value for an operator containing the spatial projection operator at that location. Whenever a particle satisfies condition (ii), it also satisfies this condition, as its forwards and backwards-travelling states always overlap. However, this can only be a necessary condition for particle presence in the sense of (i) to (vi) above, rather than a sufficient condition, as there are cases where:

\renewcommand{\labelenumi}{\arabic{enumi}).}
\begin{enumerate}
    \item A particle has a non-zero weak value of an operator containing the spatial projection operator at a location, but has no continuous path to/from this location.
    \item A particle has a non-zero weak value of an operator containing the spatial projection operator at a location, but a weak value of zero for an operator containing the spatial projection operator at a coarse-graining of that location (e.g. having a non-zero weak value of an operator containing the spatial projection operator on path $B$, and a non-zero weak value of an operator containing the spatial projection operator on path $C$, but a weak value of zero for an operator containing the spatial projection operator for the space composed of paths $B$ and $C$).
    \item A particle has a weak value of zero for the spatial projection operator along a certain path, yet has a non-zero weak value for some other operator (e.g. the spin operator in a given direction) multiplied by the spatial projection operator along that path (e.g. in the quantum Cheshire cat set-up).
\end{enumerate}

Using the two-state vector formalism and weak value tools, we can quantitatively analyse the nested interferometer set-up to show Point (2). We first define the forwards-travelling initial vector and backwards-travelling final vector by the paths via which they evolve:
\begin{equation}
    \begin{split}
        \ket{\psi_i}=\frac{\sqrt{2}\ket{A}+\ket{B}+\ket{C}}{2}\\
        \bra{\psi_f}=\frac{\sqrt{2}\bra{A}+\bra{B}-\bra{C}}{2}
    \end{split}
\end{equation}

Using these, and defining the spatial projection operator for arm $A$ as $\hat{P}_A=\proj{A}$ (similarly for $B$ and $C$), we get the same weak values as in \cite{Vaidman2013Past}:
\begin{equation}
\begin{split}
    \av{\hat{P}_A}_w=&\frac{\bra{\psi_f}\proj{A}\ket{\psi_i}}{\braket{\psi_f}{\psi_i}}=1\\
    \av{\hat{P}_B}_w=&\frac{\bra{\psi_f}\proj{B}\ket{\psi_i}}{\braket{\psi_f}{\psi_i}}=\frac{1}{2}\\
    \av{\hat{P}_C}_w=&\frac{\bra{\psi_f}\proj{C}\ket{\psi_i}}{\braket{\psi_f}{\psi_i}}=-\frac{1}{2}
\end{split}
\end{equation}

However, if we want to see if the particle was present in the inner interferometer as a whole (as in either arm $B$ or $C$), we define $\hat{P}_{BC}=\proj{B}+\proj{C}$, as we are allowed to do since projectors in standard quantum mechanics are linear. We find
\begin{equation}
\begin{split}
 \av{\hat{P}_{BC}}_w&=\frac{\bra{\psi_f}(\proj{B}+\proj{C})\ket{\psi_i}}{\braket{\psi_f}{\psi_i}}\\
 &=\frac{1}{2}-\frac{1}{2}=0
 \end{split}
\end{equation}

If we assumed a non-zero weak trace implies particle presence, this would mean the photon was never in the inner interferometer (made up of arms $B$ or $C$) overall. (\cite{Vaidman1996WeakElReality,Ben2017Improved} explicitly say that weak values obey the sum rule, and so allows us to say $\av{\hat{P}_{BC}}_w$ must equal $\av{\hat{P}_{B}}_w+\av{\hat{P}_{C}}_w$.) Taken separately the summands are taken to mean that the particle was in arm B, and was in arm C, respectively, yet the sum being $0$ is taken to mean that the particle is not in their union (i.e., not in the inner interferometer at all); this seems incoherent. (Aharonov et al consider a similar scenario in their three-box experiment, and also discuss this idea of negative weak value in the equivalent of arm C cancelling the positive weak value in the equivalent of arm B \cite{Aharonov2017Disappearing}. This usefulness of considering the sign of the weak value for resolving paradoxes is also discussed in \cite{waegell2022quantum}.) This illustrates the importance of the sign of the weak values, which the Weak Trace Approach, and the graphical form of the TSVF (as in Fig.\ref{fig:TSVFDiag}), neglect.

For weak value experiments aiming to distinguish which-path information, we can link this to the Visibility-Distinguishability Inequality \cite{Jaeger1995Complementarity,Englert1996Fringe},
\begin{equation}\label{Eq:DistVis}
    \mathcal{D}^2+\mathcal{V}^2\leq1
\end{equation}
where $\mathcal{D}$ is the distinguishability of which path the light travelled, and $\mathcal{V}$ is the fringe-visibility at the output of the interferometer. Therefore, doing anything which would increase the distinguishability between the two paths (e.g. placing different tags on $B$ and $C$) will affect the interference pattern at the $BCE$ beamsplitter. Given perfect interference is required to ensure all light that enters the inner interferometer from $D$ exits into $D3$, anything causing distinguishability between paths $B$ and $C$ will allow light to leak through onto $E$, and cause a trace to show in $D2$. Therefore, it will never be showing what would happen in an unperturbed system.

Given Eq. \ref{Eq:DistVis}, $\av{\hat{P}_{BC}}_w$ is a far better measure of whether light reaching $D2$ was ever in the undisturbed inner interferometer. This is as measuring $\av{\hat{P}_{BC}}_w$ does not cause distinguishability between paths $B$ and $C$ in the inner interferometer, and so does not affect the interference pattern required to output all inner interferometer light into $D3$. $\av{\hat{P}_{BC}}_w$  being 0 provides support for a `common-sense' path (i.e. light only travelling via $A$ to reach $D2$) in an unperturbed system.

An argument against looking at the weak value of the projector over multiple paths (e.g. $\hat{P}_{BC}$) specific to the Weak Trace Approach is that the weak trace is only defined as the effect of the particle on the environment local to it---and combining the two weak values for these two paths is not a local operation, so does not indicate anything about the weak trace. However, this still goes against classical ideas of presence, given it shows that the presence in each of the two arms must have opposite sign, so they can cancel when added, and so implies that being realist about weak values implying presence requires being realist in some way about negative presence. This negative presence is yet another way the kind of presence that is attributed by weak value approaches differs from classical presence.

Point (3) seems to support the Weak Trace Approach to particle presence, whereby a non-zero weak value of any operator corresponding to a property of the particle, multiplied by the spatial projection operator along a path, implies the particle was present along that path. However, as we saw in Section \ref{Vaidman}, taking this to imply presence even when the weak value of the spatial projection operator alone is zero, contradicts the argument in \cite{Peleg2018CommentRevisited} that the weak trace comes from the necessarily nonzero interaction between a particle and its environment, given if this were the case, we would expect a particle to necessarily have a nonzero weak value for the spatial projection operator for any path along which it travels. This is a dilemma for weak value approaches---they either contradict the rationale for weak values indicating particle presence, or imply that a particle's properties are not necessarily in the same location as that particle.

From these three points, we see that the ``nonzero weak value for an operator containing the spatial projection operator'' condition can be satisfied while the conditions for a particle being present at a location are not satisfied. Therefore, it is not a sufficient condition for particle presence in the standard sense of the term.

\section{Weak values are only defined over ensembles}

It is important to note that the weak value approach to particle presence is used to attribute presence to individual particles. This is despite the experimental results of measurements of weak values necessarily being produced by ensembles, because weak values are obtained by post-selection. It is non-obvious that we can go from facts about ensembles to facts about individual elements of that ensemble. As an example, we see that one can have a weak value of the velocity of electrons in a given set-up that is greater than the speed of light in a vacuum \cite{Rohrlich2002Cherenkov,Solli2004Superluminal,Ahnert2004Superluminal}. However, this does not necessarily mean than any of the individual electrons in the ensemble used to obtain this weak value travelled at superluminal speeds---as otherwise this would be a violation of special relativity. Instead, similar to how group velocities of wavepackets can be superluminal despite the phase velocity (or velocity of the components) being below $c$, this superluminal weak value for speed is a fact only about the ensemble, rather than any of its constituents. Similarly, there is no reason to infer facts about the presence of individual quantum particles from weak values of operators containing the spatial projection operator.

Note, recent work has shown anomalous weak values can be observed using single-particle detection \cite{Rebufello2021SinglePhoton}---however, this requires the pre- and postselected states to be protected, which effectively involves re-initialising the preselected state on the particle after coupling. This process is could be described as still considering an ensemble of results, just embodied on the same particle at different times, rather than different particles at the same time.

\section{Why adopt weak value approaches to particle presence?}

\begin{figure}
    \centering
    \includegraphics[width=\linewidth]{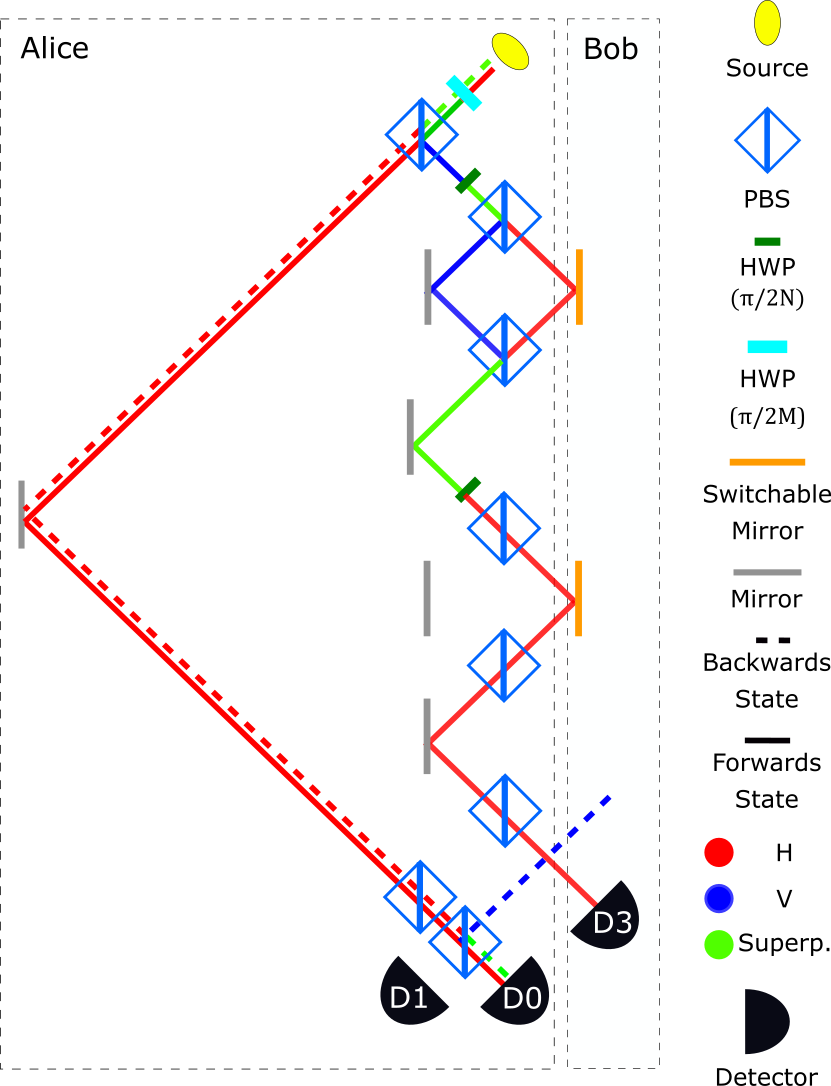}
    \caption{Two-State Vector Formalism analysis of Salih et al's polarisation-based single-outer-cycle protocol for counterfactual communication. Bob communicates with Alice by turning off/on his switchable mirrors to determine whether the photon goes to $D1$ or $D0$ respectively. We specify the polarisation, given it determines direction of travel through the polarising beamsplitters (PBSs). The forwards- and backwards-travelling states do not overlap anywhere on the inner interferometer chain when there is a detection at $D0$, meaning, by weak value approaches, the particle detected at $D0$ was never at Bob. This means Bob's ability to communicate with Alice is not explained by weak value approaches any more than it is by standard quantum mechanics.}
    \label{fig:Salih}
\end{figure}

Weak value approaches are intended to provide more (interpretational) information about the underlying state of the system than standard quantum mechanics. Vaidman \cite{Vaidman2013Past} says that an issue with Wheeler's ``common-sense" approach to particle trajectories \cite{Wheeler1978Past}, is that it is entirely operational: not telling us anything about the underlying mechanisms at work, just the final result. Similarly, \cite{Vaidman2013Past} says ``the von Neumann description of the particle alone is not sufficient to explain the weak trace''---implying weak value approaches provide something more than the standard von Neumann approach does. We however turn this criticism back on weak value approaches -- they do not tell us anything about the underlying system either, beyond standard quantum mechanics. 

A key motivation for weak value approaches is to explain how the results of interference are affected by changes to disconnected regions. They are intended to show that phenomena such as Wheeler's Delayed Choice, or Salih et al's Counterfactual Communication protocol \cite{Salih2013Protocol,Hance2021Quantum} and related effects \cite{Salih2016Qubit,Salih2018Paradox,Hance2021CFGI,Salih2021EFQubit,Salih2020DetTele} are not as ``spooky" as they appear \cite{Vaidman2014SalihCommProtocol,Vaidman2015Counterfactuality,Vaidman2016Counterfactual,Vaidman2019Analysis}. Salih et al however have shown that in their communication protocol, the weak value of the spatial projection operator (or any compound operator including the spatial projection operator) at Bob is always zero when Alice receives the quantum particle (see Fig.\ref{fig:Salih}) \cite{Salih2018Laws}. Aharonov and Vaidman have also given an altered protocol for counterfactual communication where the weak value of the spatial projection operator (or any compound operator including the spatial projection operator) at Bob is zero \cite{Aharonov2019Modification,Wander2021ApprCF}. Each of these results show that weak value approaches do not explain this phenomenon.

While suggesting a time-symmetry to quantum processes through the TSVF, weak value approaches do not imply any new physics beyond standard quantum theory \cite{Vaidman1996WeakElReality}, to explain the causes of counterintuitive quantum effects when there is a nonzero weak value of the spatial projection operator. These approaches simply assume the particle was present wherever the weak value of an operator containing the spatial projection operator is nonzero. Hence they posit particle presence but contribute nothing testable to our physics. This label confuses matters by oversimplifying a complex concept: what it means for a specific particle to be sequentially ``present" at a two specific places in quantum field theory. It also causes a number of paradoxes by itself, such as the discontinuous photon trajectories in the nested interferometer set-up above.

\section{Conclusion}

We have investigated four issues with both the Weak Trace Approach, and other approaches where a nonzero weak value of the spatial projection operator is implicitly assumed to indicate the presence of an individual quantum particle. First, we have shown even weak measurements disturb a system, so any approach relying on such a perturbation to determine the location of a quantum particle will only describe this disturbed system, not a hypothetical undisturbed state. Secondly, we have shown even assuming no disturbance, there is no reason to associate the non-zero weak value of an operator containing the spatial projection operator with the classical idea of `particle presence'. Thirdly, we have discussed that weak values are only measurable over ensembles, and so to infer properties of individual particles from values of them is problematic at best. Finally, we have shown that weak value approaches to the path of a particle do not contribute any new physics---the assumption of a connection between particle presence and weak values does not give us anything testable. This is not to say that weak values are not useful in general---as discussed, they are very useful in metrology \cite{Dixon2009WeakAmp,Jordan2014AdvWeak,Harris2017Weak}---but that these approaches specifically are not useful for helping identify the past path of quantum particles.

\textit{Acknowledgements -} We thank Hatim Salih and Sophie Inman for useful discussions, and an anonymous reviewer who gave detailed and useful feedback on multiple earlier versions of this paper. This work was supported by the University of York's EPSRC DTP grant EP/R513386/1, and the Quantum Communications Hub funded by EPSRC grants EP/M013472/1 and EP/T001011/1.

\bibliographystyle{unsrturl}
\bibliography{ref.bib}

\end{document}